\documentclass{article}
\usepackage[T1]{fontenc}

\usepackage{microtype}
\usepackage{graphicx}
\usepackage{tikz}
\usepackage{pifont}
\usetikzlibrary{arrows.meta,patterns}
\usepackage{subfigure}
\usepackage{booktabs} 
\usepackage{amssymb}
\usepackage{xspace}
\usepackage{xcolor}
\newcommand{\name}{TideSand\xspace}
\definecolor{darkkhaki}{RGB}{189,183,107}
\definecolor{teal}{RGB}{0,128,128}

\newenvironment{packeditemize}{\begin{list}{$\bullet$}{\setlength{\itemsep}{0.5pt}\addtolength{\labelwidth}{-4pt}\setlength{\leftmargin}{2ex}\setlength{\listparindent}{\parindent}\setlength{\parsep}{1pt}\setlength{\topsep}{2pt}}}{\end{list}}

\newcommand{\mypara}[1]{\par\vspace{0.2cm}\vspace{-\parskip}\noindent{\bf #1:}~\ignorespaces}

\newcommand{\tightcaption}[1]{\vspace{-0.27cm}\caption{{\normalfont{\textit{{#1}}}}}\vspace{-0.2cm}}
\newcommand{\tightsection}[1]{\vspace{-0.25cm}\section{#1}\vspace{-0.15cm}}
\newcommand{\tightsubsection}[1]{\vspace{-0.35cm}\subsection{#1}\vspace{-0.08cm}}

\usepackage{array}
\usepackage{colortbl}

\usepackage{xcolor}

\usepackage{xurl}
\usepackage{hyperref}

\usepackage{silence}
\usepackage[accepted]{mlsys2025}
\makeatletter
\renewcommand{\mlsys@appearing}{}
\renewcommand{\printAffiliationsAndNotice}[1]{%
  \stepcounter{@affiliationcounter}%
  {\let\thefootnote\relax\footnotetext{\hspace*{-\footnotesep}#1%
    \forloop{@affilnum}{1}{\value{@affilnum} < \value{@affiliationcounter}}{%
      \textsuperscript{\arabic{@affilnum}}\csname @affilname\the@affilnum\endcsname. }}}}
\makeatother

\AddToShipoutPictureFG{%
  \AtPageLowerLeft{%
    \put(0,36){\makebox[\paperwidth]{\normalfont\small\thepage}}%
  }%
}

\mlsystitlerunning{TideSand: Performance-Accurate Sandbox Multiplexing for LLM Agents}

\begin{document}

\twocolumn[
\mlsystitle{TideSand: Performance-Accurate Sandbox Multiplexing \\ for LLM Agents}



\mlsyssetsymbol{equal}{*}

\begin{mlsysauthorlist}
\mlsysauthor{Shulu Li}{equal,ucb}
\mlsysauthor{Hanchen Li}{equal,ucb}
\mlsysauthor{Xiaokun Chen}{harbor}
\mlsysauthor{Qiuyang Mang}{ucb}
\mlsysauthor{Hanfei Chen}{gatech}
\mlsysauthor{Jacky Kwok}{stanford}
\mlsysauthor{Hanfei Yu}{ucb}
\mlsysauthor{Alvin Cheung}{ucb}
\mlsysauthor{Edward A.~Lee}{ucb}
\mlsysauthor{Joseph E.~Gonzalez}{ucb}
\mlsysauthor{Ion Stoica}{ucb}
\end{mlsysauthorlist}

\mlsysaffiliation{ucb}{University of California, Berkeley}
\mlsysaffiliation{stanford}{Stanford University}
\mlsysaffiliation{harbor}{Harbor Framework}

\mlsysaffiliation{gatech}{Georgia Institute of Technology}

\mlsyskeywords{Machine Learning, MLSys}

\vskip 0.3in 

\begin{abstract}

As large language model (LLM) agents are deployed for increasingly long-horizon and complex tasks such as software engineering, performance tuning, and auto research, the demand for sandboxes that provide isolated execution environments for them grows rapidly. 
A typical sandbox allocates a fixed amount of resources for the lifetime of each agent, even when an LLM agent is waiting for inference, during which the sandbox is mostly idle. This causes resource utilization to be low. 
Prior work leverages statistical multiplexing by overcommitting resources to exceed the host's capacity since not all sandboxes are active at the same time.
However, several challenges arise when the host resource is multiplexed to multiple sandboxes. 
Most importantly, agent sandboxes are bursty, and when bursts of multiple sandboxes overlap, contention can leave sandboxes with fewer resources than promised, slowing down tool calls.
This compromises tool-call fidelity, which we define as sandboxes' ability to preserve tool-call execution times and results.
For emerging agent workloads that rely on accurate performance measurements, sandbox systems need to maximize resource utilization while preserving tool-call fidelity to better serve the workload.

We propose \name, a sandbox system that intelligently multiplexes resources for performance-sensitive LLM agents, such as auto research agents working on system measurement tasks. 
\name combines three key ideas: agent-aware scheduling, which schedules sandbox execution intelligently at the tool call level to avoid resource contention;
speculative swapping, which offloads the memory of sandboxes waiting on inference and restores it before their next tool call;
and selective rollback, which checkpoints selected sandboxes when contention occurs and restores them when machines are idle. 
We evaluate \name across six model-workload pairs on AlgoTune, Terminal-Bench 4.0, and Terminal-Bench Science. \name achieves 3.8--10$\times$ higher throughput than strict reservation while maintaining tool-call fidelity, or 7--44\% higher fidelity than overcommit baselines at similar throughput. Moreover, \name reduces the evaluation error on Algotune by 19$\times$ through improved tool-call fidelity. We hope the work motivates future research to maintain tool call fidelity while keeping resource utilization high through statistical multiplexing. We open-source our code at \href{https://github.com/TideSand/tidelet/tree/paper}{https://github.com/TideSand/tidelet}.
\end{abstract}

]



\printAffiliationsAndNotice{\textsuperscript{*}Equal contribution. }

\section{Introduction}
\label{sec:introduction} 

The cost of serving large language model (LLM) agents extends beyond inference. Agents also need execution environments in which to edit code,
 run tests, analyze data, and conduct experiments~\cite{sweagent,openhands,scienceagentbench}.
 These environments, or \emph{sandboxes}, isolate tool execution and retain files, processes, and memory state across successive calls~\cite{DSec,DeltaBox}.
 Unlike an individual tool call, a sandbox can remain allocated for an entire agent task, including intervals spent waiting for model inference~\cite{DSec,AgentZip}.
As inference costs at a given capability have fallen in recent years~\cite{aiindex2025},
the CPU and memory for these sandboxes can become a growing fraction of the total cost.
Reducing this cost requires adapting sandbox management to the workflow of the agents.

Agent sandboxes are unique in their workload characteristics: 
tool execution is often bursty. For example, a compilation task might require several CPUs to run, and then wait for an LLM inference call to process its output. To facilitate the requirements of the compilation phase, a sandbox is often allocated a fixed set of resources during its lifetime. However, this causes resource utilization to be low, since resources remains allocated and sits idle while the agent waits for inference. 
A natural solution to improve utilization is to leverage \emph{statistical multiplexing} (i.e., not all environments are active at the same time) 
by over-subscribing resources. Assume a trivial example in which we have two agent sandboxes. Both sandboxes require one CPU every other second, and spend the idle second waiting for remote LLM inference. If we allocate one CPU per sandbox, the utilization is 50\%. Instead, we can multiplex two sandboxes and run them on a single CPU, cutting the resource need by half.
Current systems such as DeepSeek Elastic Compute (DSec)~\cite{DSec} and AgentZip~\cite{AgentZip} improve sandbox density through this multiplexing approach.

However, resource multiplexing raises several challenges. 
First, admission control. If not enough resources are available for the sandbox to resume, we might need to delay its execution until some resources are freed to prevent thrashing. 
Second, each sandbox has state associated with it. 
In order for another sandbox to utilize the same resource, we must first offload its state and, before resuming it, 
restore its state. Yet the overhead associated with offload and restore might negate the advantages of multiplexing, 
so we need to choose carefully when to do so.

\begin{figure}[t]
    \centering
    \includegraphics[width=\columnwidth]{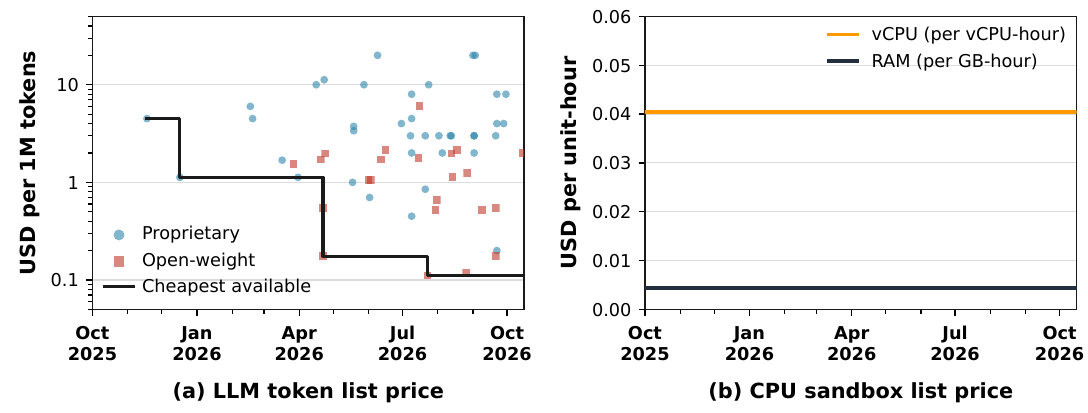}

    \tightcaption{Equal-Intelligence LLM Token prices plummet while CPU sandboxes cost stay relatively flat over the last year.}
    \label{fig:llm_vs_sandbox_prices}
\end{figure}

Most importantly, in many scenarios, agents aim to improve the performance of a system. 
In this case, accurate resource provision and measurement is crucial, as this is the signal the agent depends on to iterate. 
Consider a profiling task that requests ten CPU cores but receives only six for most of its execution.
The tool call response may reflect interference from other sandboxes rather than the behavior of the implementation.
 An agent acting on this report will mistake a change in resource availability for a change in program performance.

To measure this effect, we use \emph{tool-call fidelity} to describe how well the execution timing and results are preserved when executing in a multiplexed sandbox. Practically, we run the same tool call in a resource guaranteed sandbox, and again in a multiplexed sandbox, and compare their tool call timing and results. A sandbox system with high tool-call fidelity should have similar results between the two. 

A unique opportunity here is to schedule the agent tool-calls intelligently. Delaying the LLM response slightly has no effect on the agent, since this is indistinguishable from slower LLM inference. This allows a scheduler to defer tool-calls until sufficient resources are available, while being totally transparent to the sandbox. This adds a queuing delay, but preserves tool-call fidelity.

This requirement leads to an objective of maximizing \emph{sandbox goodput}: the rate of successfully completed agent tasks whose tool calls satisfy a specified fidelity requirement.
 We assess fidelity by comparing tool outcomes and execution-time distributions against a reference with the configured resources and no contention from other sandboxes. 
 Using distributions accounts for the variation present even in uncontended executions. 
 Goodput therefore captures the benefit of serving more agents together with the cost of retries, failures, and distorted observations.

To maximize goodput, we propose \name, which oversubscribes sandbox resources while protecting tool-call fidelity. Using agent-loop signals together with resource observations, it applies three mechanisms:
\begin{packeditemize}
    \item \textbf{Agent-aware Scheduling} admits additional work only when sufficient resource headroom is available in CPU, memory, and bandwidth, reducing resource thrashing and thus preserving tool call fidelity. 
    \item \textbf{Speculative Swapping} moves memory in idle sandboxes to disk during inference intervals when there is no active tool execution. When a tool call is about to be executed, it restores memory from swap before so that reclaiming memory does not simply shift recovery costs into tool execution. 
    \item \textbf{Selective Rollback} handles sudden spikes in resource requirements and residual contention by selectively check-pointing sandboxes before tool call execution, and shutting down sandboxes when resources are saturated to release headroom. These tool calls are re-executed after resources are available for execution.
\end{packeditemize}

We evaluate \name across six model-workload pairs on AlgoTune, Terminal-Bench 4.0, and Terminal-Bench Science with Sonnet 5.5 and DeepSeek V4.1 Flash.
\name achieves 3.8--10$\times$ higher throughput than strict reservation while maintaining tool-call fidelity, or 7--44\% higher fidelity than baselines including overcommit at similar throughput.
Moreover, by improving tool-call fidelity, \name reduces the evaluation error on AlgoTune, measured against the agent's task reward under strict reservation by 19$\times$.

This paper makes three contributions:
\begin{packeditemize}
    \item We define \emph{tool-call fidelity} as a metric for capturing resource fairness in agent sandboxes, and formulate \emph{goodput} as the overarching goal for sandbox systems for serving agents.
    \item We design \name, a sandbox system that combines agent-aware scheduling, speculative swapping, and selective rollback to exploit idle resources while protecting tool-call fidelity.
    \item We show \name achieves 3.8--10$\times$ higher throughput than strict reservation or 7--44\% higher fidelity than baselines at similar throughput. It also reduces the task-reward error by 19$\times$ from fidelity improvement.
\end{packeditemize}

\section{Background}
\label{sec:background}
\subsection{Emerging Agent Workloads}
LLM agents perform tasks in software engineering~\cite{sweagent,openhands}, scientific data analysis~\cite{scienceagentbench}, and web automation~\cite{webarena}.
Platforms such as OpenHands execute tools for editing files, running tests, and driving a browser inside an isolated execution environment known as a \emph{sandbox}~\cite{openhands}.
Because an agent task unfolds over multiple turns where later actions build on earlier ones, a stateful sandbox retains filesystem, process, and memory state across the trajectory~\cite{DSec,DeltaBox}.

More recently, people have explored using agents to conduct autoresearch~\cite{frontiercs, frontiercs2, algotune}. 
In these workflows, agents will usually be given a research environment and an end goal.
The agent will conduct profiling, analysis, and tuning to come up with innovative solutions to more practical and open-ended tasks.
Notably, because these tasks target practical systems such as databases, they require substantially more resources.



\subsection{Increasing Sandbox Cost}
To serve these agents, systems need to consider both LLM inference and sandbox costs.
And sandbox costs are surprisingly becoming more and more important.
Figure~\ref{fig:llm_vs_sandbox_prices} compares October 2026 listed token prices for models scoring at least 50\% on Terminal-Bench 2.1~\cite{terminalbench21}, 
grouped by release date. 
The minimum plotted price falls from \$4.50 for the November 2025 release cohort to \$0.111 per million tokens for July 2026 releases, a roughly 40$\times$ difference (assuming a 3:1 input/output blend as in Table~\ref{tab:llm_prices}). 
For reference, AWS Fargate's Linux/x86 on-demand rates in US-east-1 stay stable on the right.

\begin{figure}[t]
    \centering
    \includegraphics[width=\columnwidth]{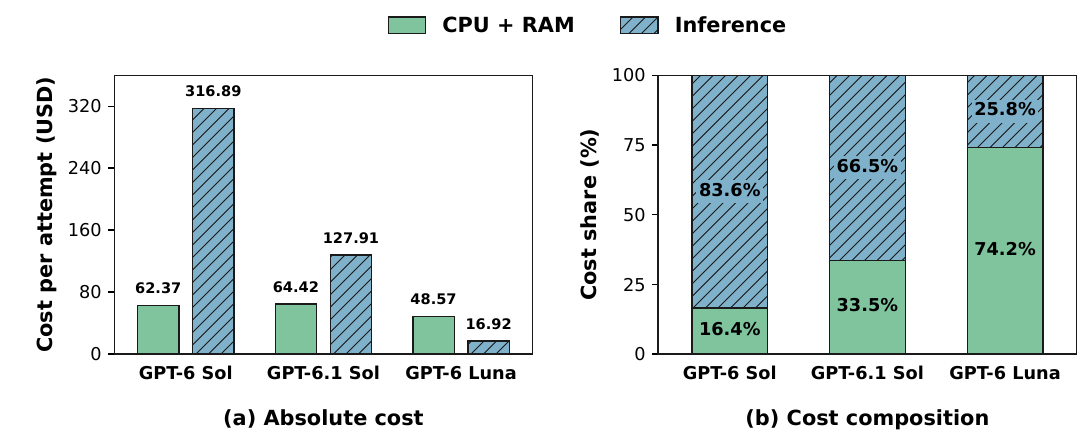}
    \vspace{-15pt}
    \tightcaption{End-to-end agent cost breakdown across models at max effort for Terminal-Bench 4.0, excluding GPU tasks}
    \label{fig:agent_cost_breakdown}
\end{figure}

Moreover, sandboxes already account for a substantial share of agent serving costs.
Figure~\ref{fig:agent_cost_breakdown} shows CPU and memory accounting for 16.4\%, 33.5\%, and 74.2\% of total cost for GPT-6 Sol, GPT-6.1 Sol, and GPT-6 Luna respectively when we run them at max effort for Terminal-Bench 4.0 even after excluding tasks requiring GPUs.
This staggering comparison, especially how models are getting increasingly cheap even for the same model type, demonstrates sandbox could soon become the biggest bottleneck in agent serving.

\subsection{Prior Work on Agent Sandboxes}

\mypara{System Mechanisms}
Firecracker~\cite{Firecracker} and gVisor~\cite{gVisor} isolate execution through lightweight microVMs and a userspace application kernel, respectively.
TrEnv-X~\cite{TrEnv-X} reduces startup and memory costs through reusable sandboxes and memory templates backed by CXL/RDMA pools.
DeltaBox~\cite{DeltaBox} accelerates checkpoint and rollback through incremental state capture and frozen process templates, overlapping checkpoint work with inference.
These mechanisms provide building blocks for efficient sandbox scheduling.

\mypara{Workload-Aware Optimizations}
Recent systems increasingly tailor these mechanisms to agent workloads, often using oversubscription to share resources left idle between tool calls.
DeepSeek Elastic Compute (DSec)~\cite{DSec} combines oversubscription, memory reclamation, and RL-coordinated sandbox lifecycles, while CPU priorities and core scheduling protect latency-sensitive work.
SpecBox~\cite{SpecBox} predicts sandbox requests from streaming model output and tool transitions to overlap preparation with ongoing inference or execution.
AgentZip~\cite{AgentZip} compresses memory using template and cross-sandbox redundancy, schedules compression during inference, and prefetches pages to reduce restoration stalls.
AgentCgroup~\cite{AgentCgroup} handles contention in oversubscription by throttling competing tool calls,
but it comes at the cost of slowing down tool calls for too long.

DSec, SpecBox, AgentZip, and AgentCgroup adapt resource sharing and oversubscription to agent workloads, while prior works provide valuable primitives for sandbox scheduling.
\name\ builds on the system primitives and invents new workload-aware scheduling policies to better support emerging performance sensitive agent workloads.

\begin{table}[t]
    \centering
    \tightcaption{Sandbox mechanisms and objectives: \checkmark\ denotes an explicit focus; \textendash\ denotes otherwise.}
    \label{tab:functionality}
    \vspace{0.32cm}
    \definecolor{sandrow}{RGB}{226,240,240}
    \footnotesize
    \setlength{\aboverulesep}{0pt}
    \setlength{\belowrulesep}{0pt}
    \setlength{\tabcolsep}{0pt}
    \setlength{\extrarowheight}{2.5pt}
    \renewcommand{\arraystretch}{1.15}
    \begin{tabular}{@{}>{\raggedright\arraybackslash}m{0.22\columnwidth}*{4}{>{\centering\arraybackslash}m{0.195\columnwidth}}@{}}
        \toprule
        System & Oversub-\newline scription & Memory\newline reclamation & Inference-\newline aware & Fidelity-\newline preserving \\
        \midrule
        DSec & \checkmark & \checkmark & \textendash & \textendash \\
        SpecBox & \textendash & \textendash & \checkmark & \textendash \\
        AgentZip & \checkmark & \checkmark & \checkmark & \textendash \\
        AgentCgroup & \checkmark & \textendash & \checkmark & \textendash \\
        \midrule
        \rowcolor{sandrow}
        \textbf{\name} & \checkmark & \checkmark & \checkmark & \checkmark \\
        \bottomrule
    \end{tabular}
\end{table}

\section{Sandbox for Performance Sensitive Agents}
\label{sec:motivation}
While prior work in sandbox overcommitting increases resource utilization, 
multiplexing alone is insufficient for providing performance-accurate sandboxes.

\begin{figure}[t]
    \centering
    \includegraphics[width=\columnwidth]{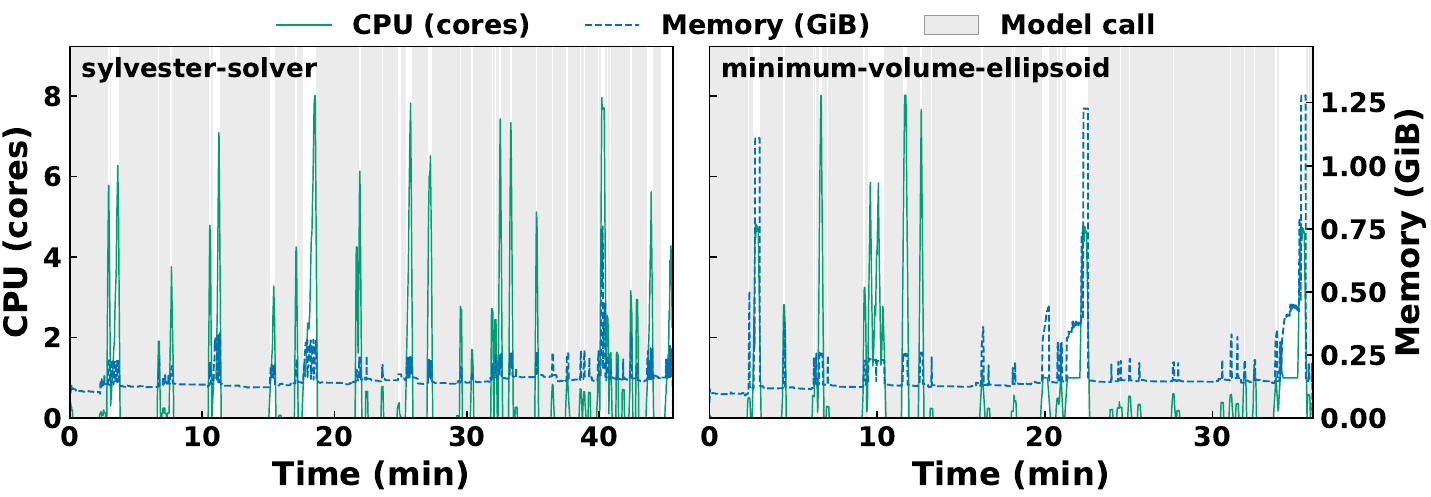}
    \vspace{-10pt}
    \tightcaption{Measured CPU usage (cores) and memory usage (GiB) over time for two tasks. Shaded intervals indicate model calls. CPU usage is bursty, while memory remains resident between execution bursts.}
    \label{fig:workload-burstiness}
\end{figure}

\tightsubsection{Workload Characteristics and Multiplexing}
\label{sec:workload-characteristics-and-oversubscription}
Agent sandboxes have bursty demand for CPU and memory. Tool calls can require several CPU cores,
while inference leaves those cores largely idle and memory resident
(Figure~\ref{fig:workload-burstiness}). Provisioning each sandbox for peak demand throughout a task wastes capacity that could serve other agents.
When tool execution is staggered across agents, a host can oversubscribe: accommodate sandboxes whose aggregate configured resources exceed its physical capacity.

For example, DSec~\cite{DSec} overcommits by admitting a sandbox when the node still has measured CPU and memory capacity, 
so the sandboxes' configured limits can sum to more than the host's physical capacity.

\subsection{Resource Contention under Multiplexing}
Multiplexing faces challenges in predicting future demand. 
Several admitted agents may finish inference at the same time and launch resource-intensive tools, creating contention even when average host utilization is moderate.
For example, DSec checks local resource capacity before creating a sandbox and combines CPU priorities with core scheduling to limit interference~\cite{DSec}.
These controls operate at sandbox-level granularity, which cannot ensure that resources are ready for these bursts.

\begin{figure}[t]
    \centering
    \includegraphics[width=0.8\columnwidth]{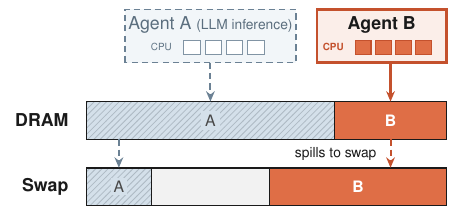}
    \tightcaption{Under naive multiplexing, agent A not yet doing a tool call is using sandbox resources from active agent B.}
    \label{fig:mem-swap}
\end{figure}

On the other hand, the agent loop offers an opportunity for speculative preparation. 
As a tool call streams, the scheduler can anticipate execution and begin restoring reclaimed memory 
while the model generates the remaining arguments. 
Moving restoration into this interval can reduce overhead for the tool call. 
Conversely, retaining inactive pages from idle sandboxes from agents doing inference can force active tools to spill to swap as shown in Figure~\ref{fig:mem-swap}.


\begin{figure}[t]
    \centering
    \subfigure[Tool-call slowdown.\label{fig:tool-fidelity-left}]{%
        \includegraphics[width=0.48\columnwidth]{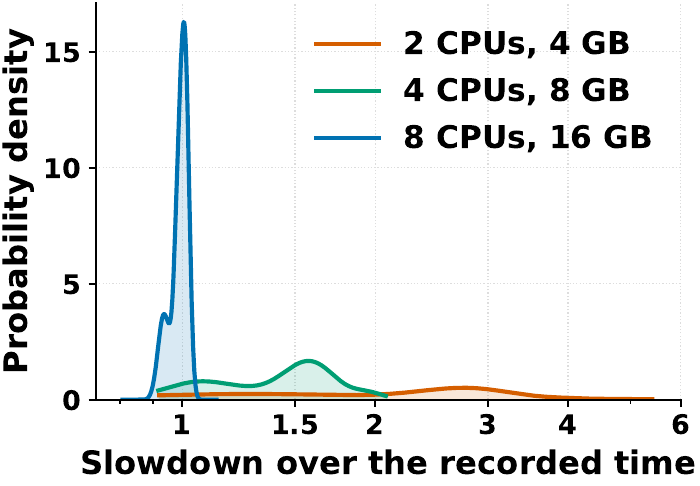}}\hfill
    \subfigure[Final benchmark reward.\label{fig:tool-fidelity-right}]{%
        \includegraphics[width=0.48\columnwidth]{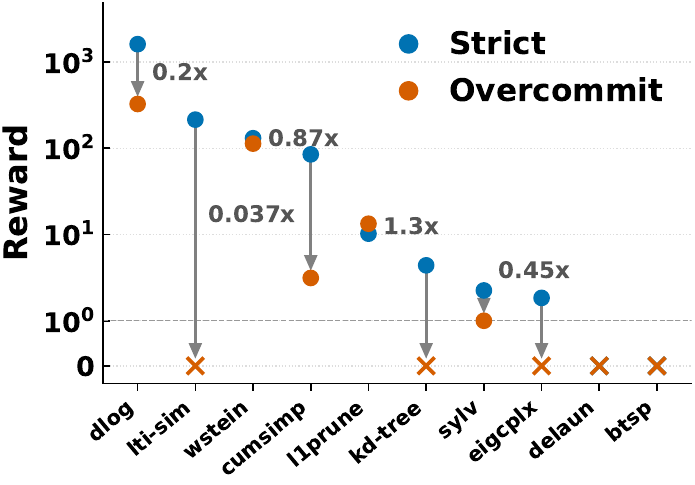}}
    \tightcaption{Resource contention causes tool execution time to deviate from the reference and affects final benchmark reward.}
    \label{fig:tool-fidelity}
    \vspace{-5pt}
\end{figure}

\subsection{Tool Call Fidelity}
Most importantly, resource contention caused by multiplexing can cause a more severe consequence for the emerging performance-sensitive sandboxes such as auto research agents.
Autoresearch agents conduct experiments and use the observations to guide further investigation~\cite{aiscientist}. 
For example, AlgoTune's agent profiles candidate programs and uses their measured runtimes to select faster correct implementations~\cite{algotune}.
If the sandbox receives fewer CPU cores or less resident memory than the experiment assumes, the observation can be distorted by resource contention.
The need to keep a tool call's result and execution time consistent with a reference run on its configured CPU and memory motivates a property of sandbox execution, which we call \emph{tool-call fidelity}. In other words, even if the call returns the correct output, it can still lose fidelity when CPU sharing or memory pressure changes its runtime.

We show the effect of resource contention on tool execution time and final benchmark reward in Figure~\ref{fig:tool-fidelity}.
The left figure compares execution-time distributions for 10 tool calls each repeated for 10 times under three CPU and memory allocations, 
normalized to the recorded time with 8 CPUs and 16\,GB. 
Reducing the allocation increases both execution time and its variation.
The right figure compares final benchmark rewards under strict reservation and a naive overcommitment policy based on resource usage.
Rewards change in both directions, showing that resource contention can change the agent's outcome even on the same task. 
A higher reward under overcommitment does not imply better fidelity: in either direction, the result differs from the reference execution.
\begin{figure*}[t]
    \centering
    \includegraphics[width=0.9\textwidth]{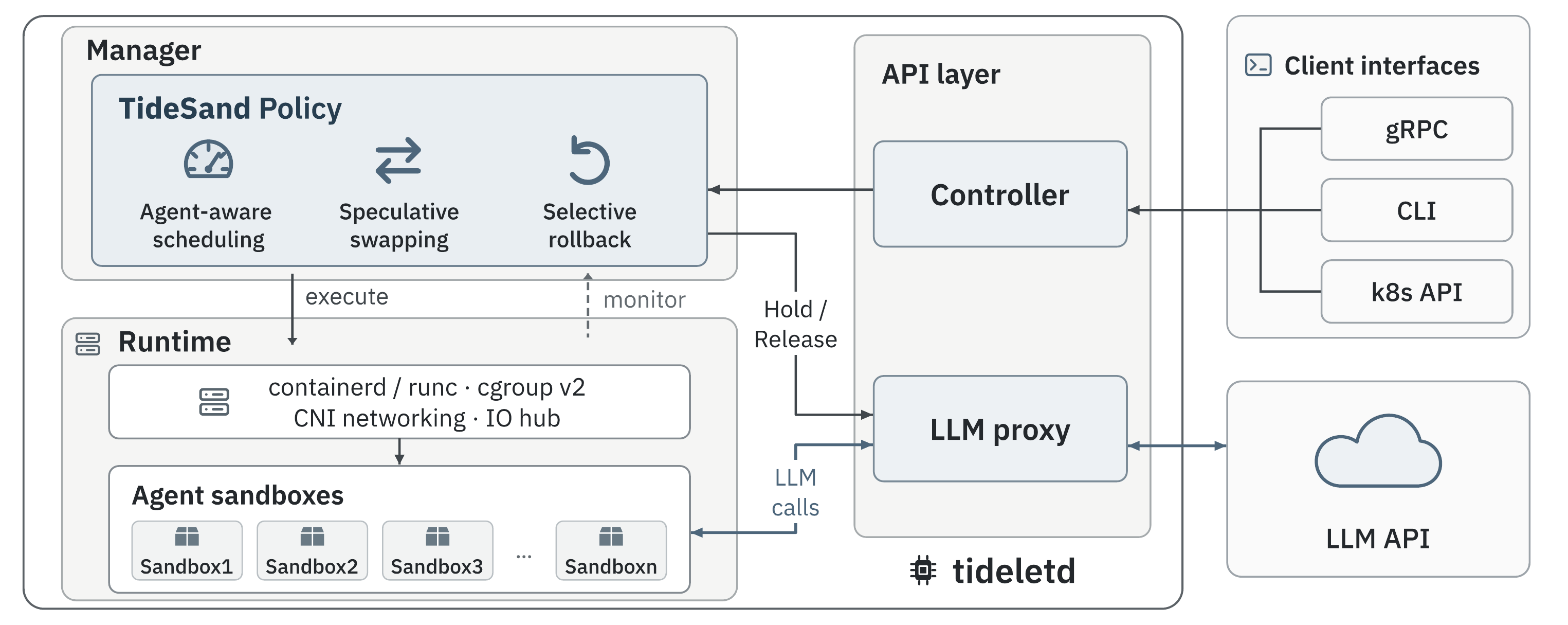}
    \vspace{-5pt}
    \tightcaption{System overview of \name.}
    \label{fig:system-overview}
    \vspace{-5pt}
\end{figure*}

Given the impact of resource contention on experimental results, tool-call fidelity is the first priority in sandbox systems for autoresearch agents.
It is therefore even preferable to delay a tool call rather than run it immediately with insufficient resources and distort its results,
since queueing delay is indistinguishable for LLM inference delay. 
On the other hand, slowdown from contention during execution becomes part of the measurement and can mislead the agent's next steps.

\subsection{Improving Sandbox Goodput}
The opportunity for statistical multiplexing therefore comes with a requirement to preserve execution conditions when those calls run.
Increasing sandbox concurrency is useful only when it does not compromise tool-call fidelity. 
We capture this goal as \emph{sandbox goodput}: the rate of successfully completed agent tasks whose tool calls satisfy a specified fidelity requirement. 
Improving goodput requires reclaiming resources during inference, restoring them before the next tool call, and controlling concurrent execution when demand exceeds host capacity. 


\section{System Design}
\label{sec:system}

\subsection{System Overview}

\name is a sandbox control plane that statistically multiplexes host resources across multiple agent sandboxes. This sharing improves utilization because sandbox demand is bursty during tool execution and typically falls to a fraction of the requested resources while agents wait for LLM responses. The key insight of \name is to schedule individual tool calls rather than solely at sandbox admission.

Figure~\ref{fig:system-overview} shows the architecture of our system.
\name runs as a policy inside \texttt{tideletd}, a host-local daemon with four components. The \emph{controller} exposes interfaces for managing sandboxes through gRPC, a CLI, and Kubernetes.
The system operates transparently to the sandbox. The only configuration change required is to direct its LLM endpoint to the \emph{LLM Proxy} managed by \texttt{tideletd}, allowing \name to observe and schedule tool calls without modifying the agent harness.
The \emph{manager} hosts the scheduling policy, which makes scheduling decisions and controls the \emph{runtime}. Our \emph{runtime} uses containers managed through cgroups, but it can be replaced by another sandbox technology, such as microVMs, provided that it supports the operations required by the \emph{manager}.

The manager comprises three components.
The \emph{resource monitor} communicates with the runtime implementation and reports per-sandbox CPU usage, memory traffic, and tool-call activity.
The \emph{scheduler} decides when to admit sandboxes, execute tool calls, swap memory, and checkpoint sandboxes. It passes these decisions to the \emph{action executor}, which issues the corresponding commands to the \emph{LLM Proxy} and the \emph{runtime}.


\subsection{Agent-aware Scheduling}
\label{sec:scheduling}

\mypara{LLM Proxy}
\name observes the agent's inference--execution loop and controls tool-call delivery through an LLM proxy.
The agent directs its model API requests to the proxy, which preserves request and response contents while controlling response timing; no changes to the agent harness are required beyond configuring the endpoint.
A model request signals a transition to inference, while a tool-bearing response identifies the next execution interval.
Before delivering that response, the proxy reports the tool names to the scheduler.
Responses containing only recognized lightweight tools, such as file reads and edits, bypass CPU admission; those containing external execution tools, such as shell commands, wait until the scheduler admits them.
For streaming responses, the proxy forwards the text preceding the first tool-call event and buffers that event and the remaining response, preventing the agent from starting a tool before admission while allowing model generation to continue.
Responses without tool calls are not held for resource admission.

\mypara{Statistical Resource Prediction}
A tool call's resource requirements are unknown before execution. High-fidelity execution requires sufficient resources to be available on the host, but reserving the sandbox's full configured allocation for every call wastes capacity when demand is lower.

To balance fidelity and utilization, \name predicts each call's resource usage before execution and reserves the predicted amount.
The prediction is learned from the sandbox's recent calls.
Let $u_i(t)$ be the measured CPU use of sandbox $i$ and $L_i$ its configured CPU limit.
During each heavy call $k$, the scheduler tracks an exponential average of $u_i(t)$ with a time constant of 10 seconds and records its peak $p_{i,k}$ when the call ends.
The predicted demand $\hat{d}_i$ of the next call is the largest peak of the last three calls; a sandbox without history uses the median peak observed on the host, or $L_i$ before any call has completed.
The reservation is
\begin{equation}
r^c_i = \left\{ \begin{array}{ll} L_i & \mbox{if } \hat{d}_i \ge L_i/2, \\  \max(\hat{d}_i, g_i) & \mbox{otherwise,} \end{array} \right.
\label{eq:reservation}
\end{equation}
where $g_i$ is the sandbox's background use between calls.
A call predicted at half its limit or more reserves the whole limit, to preserve the parallelism available to CPU-intensive jobs in the reference run.
Memory traffic is predicted the same way from each call's sustained traffic $\hat{b}_i$, with the reservation $r^b_i = \min(\hat{b}_i, B L_i / C)$ capped at the sandbox's share of the host bandwidth $B$ by its CPU limit.
A new sandbox is expected to use $\hat{m}$ bytes of memory, the median resident memory of existing sandboxes.

\mypara{Fine-grained Admission}
\name admits work at two granularities: a tool call is released when its reservation fits the host's headroom, and a sandbox is started when its expected memory fits.
Let $C$ be the host's cores and $B$ its sustainable memory bandwidth, and let $U_c$ and $U_b$ be the measured CPU use and memory traffic of all sandboxes.
Measurements can lag newly released calls. The scheduler therefore adds a pending term over calls released within the last 10 seconds, denoted by $\mathcal{R}$, accounting for the portion of each reservation not yet reflected in measured use:
\begin{equation}
P_c = \sum_{j \in \mathcal{R}} \max(0,\, r^c_j - u_j),
\end{equation}
and likewise $P_b$ for traffic.
A heavy response from sandbox $i$ is released when
\begin{equation}
\begin{array}{rcl}
U_c + P_c + r^c_i &\le& \alpha C,\\
U_b + P_b + r^b_i &\le& \beta B.
\end{array}
\label{eq:headroom-admission}
\end{equation}
with budget thresholds of $\alpha = 0.7$ and $\beta = 0.8$.
Without the pending term, a burst of responses would be released against the same apparent headroom before any of them appears in a sample.
Responses that do not fit wait in a queue ordered by arrival.
The scheduler considers the oldest response first but lets a younger response pass when it fits and the older one does not, avoiding unnecessary idling behind a large reservation.
A new sandbox starts when its boot allowance satisfies Equation~\ref{eq:headroom-admission} and its expected memory fits the admission target,
\begin{equation}
M + \hat{m} + M_{\mathrm{swapped}} \le \mu M_{\max},
\label{eq:memory-admission}
\end{equation}
with $\mu = 0.8$, where $M$ is the memory in use and $M_{\mathrm{swapped}}$ is the memory that swapped-out sandboxes with waiting responses will take back.

\subsection{Speculative Swapping}
\label{sec:swapping}

\mypara{Memory Swapping}
Memory cannot be multiplexed as readily as CPU capacity because it holds persistent sandbox state. Reusing that capacity requires offloading the state and restoring it before the sandbox accesses it again.
Offloading state to disk consumes bandwidth that active tool calls may also need. For CPU-intensive workloads, this cost may be unnecessary because CPU capacity is the bottleneck rather than memory. \name therefore initiates proactive swapping only under memory pressure.
Admission stops at 80\% of memory capacity (Equation~\ref{eq:memory-admission}), but the sandboxes already admitted keep growing.
When their growth pushes the memory in use past 90\%, the scheduler selects sandboxes waiting for model responses, prioritizing those with the largest resident working sets.
For each selected sandbox, the scheduler asks the kernel to reclaim 80\% of the sandbox's resident memory to swap, without stopping its processes.

\mypara{Speculative Prefetch}
A swapped-out sandbox must recover its working set before its next tool call to avoid swap-in page faults during execution.
The LLM proxy tells the scheduler when that moment comes: the next tool-bearing model response signals that the agent is about to resume execution.
When that response arrives at the proxy, the scheduler holds it and issues a prefetch for the sandbox.
The prefetch asks the kernel to read the sandbox's swapped pages back into memory, after which the response can proceed; heavy responses remain subject to tool-call admission, but memory swap-in is now overlapped with waiting admission.

\subsection{Selective Rollback}
\label{sec:rollback}
\label{sec:backoff}

Resource prediction is statistical: it reserves what a sandbox's recent calls indicated, not what its next call may do.
A tool call can use more resources than its predicted resource allocation, and several admitted calls can burst beyond their predictions at the same time. If the burst exceeds the headroom, the host saturates, which decreases tool-call fidelity.
A simple solution to reduce the chances of such spikes is to leave a large headroom below capacity, but doing so reduces resource utilization.
\name instead runs close to the capacity and treats saturation as a rare event to recover from rather than a state to prevent.
Selective rollback removes selected calls from execution so that the remaining calls retain sufficient resources, then retries the stopped calls once the spike has passed.

\mypara{Checkpointing Sandboxes}
A call can be undone only if its sandbox can be returned to the state before it started.
\name takes a whole-sandbox checkpoint at the moment the agent sends the model request that precedes the tool call, capturing the state of its processes and its filesystem.
Taking the checkpoint at the request rather than at the response overlaps checkpointing with model inference, reducing the delay exposed to the agent.
A rollback stops the sandbox, restores the checkpoint, and resumes the agent at the point of the request; the agent re-sends the request and the response carries the same tool call, which the scheduler now admits like any other.
However, checkpointing occupies valuable resources that could be used for running other tasks. \name therefore checkpoints only selected calls.

\mypara{Choosing Checkpoints and Restores}
A call is selected for checkpointing only when it could cause a spike and its estimated checkpoint cost is low relative to its duration: it must reserve at least half its CPU limit, the host must be loaded enough for a spike to matter, and the expected checkpoint time, estimated from earlier checkpoints, must be a small fraction of the sandbox's typical call duration.
Rollback is triggered when the measured CPU use or memory traffic has stayed above 85\% of the host's capacity for three seconds.
The scheduler then stops the checkpointed calls whose measured use exceeds their reservation by the most, until their removal is expected to bring demand below the threshold.
A stopped call is restored at the head of the queue as soon as its reservation fits.
Across the workloads in \S\ref{sec:eval}, fewer than 3\% of calls are checkpointed and fewer than 0.5\% are rolled back, with a CPU overhead under 2\% (Figure~\ref{fig:eval-micro}).

\tightsection{Implementation}
\label{sec:implementation}

\mypara{Sandbox daemon}
\name is implemented as a policy of \texttt{tideletd}, 
a host-local sandbox daemon written in about 21K lines of Go.
The daemon owns the sandbox table, runs containers through containerd and runc under cgroup v2, exposes a gRPC API and a command-line client, and can register the host as a Kubernetes node through a virtual-kubelet provider so that a Pod maps to a sandbox.
Each sandbox is created with the CPU and memory limits its task specifies; the limits bound the sandbox but are not reserved on the host.
The daemon samples every sandbox at 250\,ms intervals, recording its cumulative CPU time, resident working set, swap, and pressure-stall information, and reads the host's memory traffic from the hardware last-level-cache miss counter.
A policy is implemented behind a small event interface: it receives resource samples, model-request and model-response events from the LLM proxy, and sandbox lifecycle events, and returns actions such as release or hold a response, admit or queue a sandbox, reclaim or prefetch memory, checkpoint, roll back, and restore.
Strict reservation, threshold overcommit, and \name are policies of this interface; the \name policy is about 1,500 lines and is selected by a daemon flag.
\mypara{LLM proxy}
The daemon runs a pass-through proxy for the model API that the agent reaches through its base-URL setting, so Claude Code runs unmodified in the sandbox.
The proxy forwards streamed tokens until the response carries a tool call and buffers the remainder; it then reports the tool names to the policy and delivers the buffered part when the policy releases it.
The proxy also records every request and response, which gives the scheduler each sandbox's execution and inference intervals without instrumenting the agent.
\mypara{Memory and checkpoints}
Speculative swapping uses the kernel's reclaim and \texttt{process\_madvise} interfaces on the sandbox's processes, so swapping out and prefetch happen outside the agent.
Selective rollback uses whole-container checkpoints: the daemon freezes the sandbox's cgroup, dumps its processes with CRIU, and snapshots the writable layer of its filesystem; a rollback restores both and the agent re-issues the model request that preceded the checkpoint.
Checkpoint images live on a RAM-backed filesystem and are discarded when the call completes.
\mypara{Replay harness}
To compare policies on identical agent behavior we built TideBox, about 6K lines of Python.
It records an agent run once, capturing every model request and response with its timing together with the sandbox telemetry, and replays the recording in a fresh sandbox: 
a per-sandbox controller answers the agent's model requests from the tape through the proxy while the tool calls execute for real in the sandbox.
The replay driver paces sandbox arrivals, repeats each recording a configurable number of times, and compares every tool call's output and execution time with the recording and with reference replays.

\begin{figure*}[!t]
    \centering
    \includegraphics[width=\textwidth]{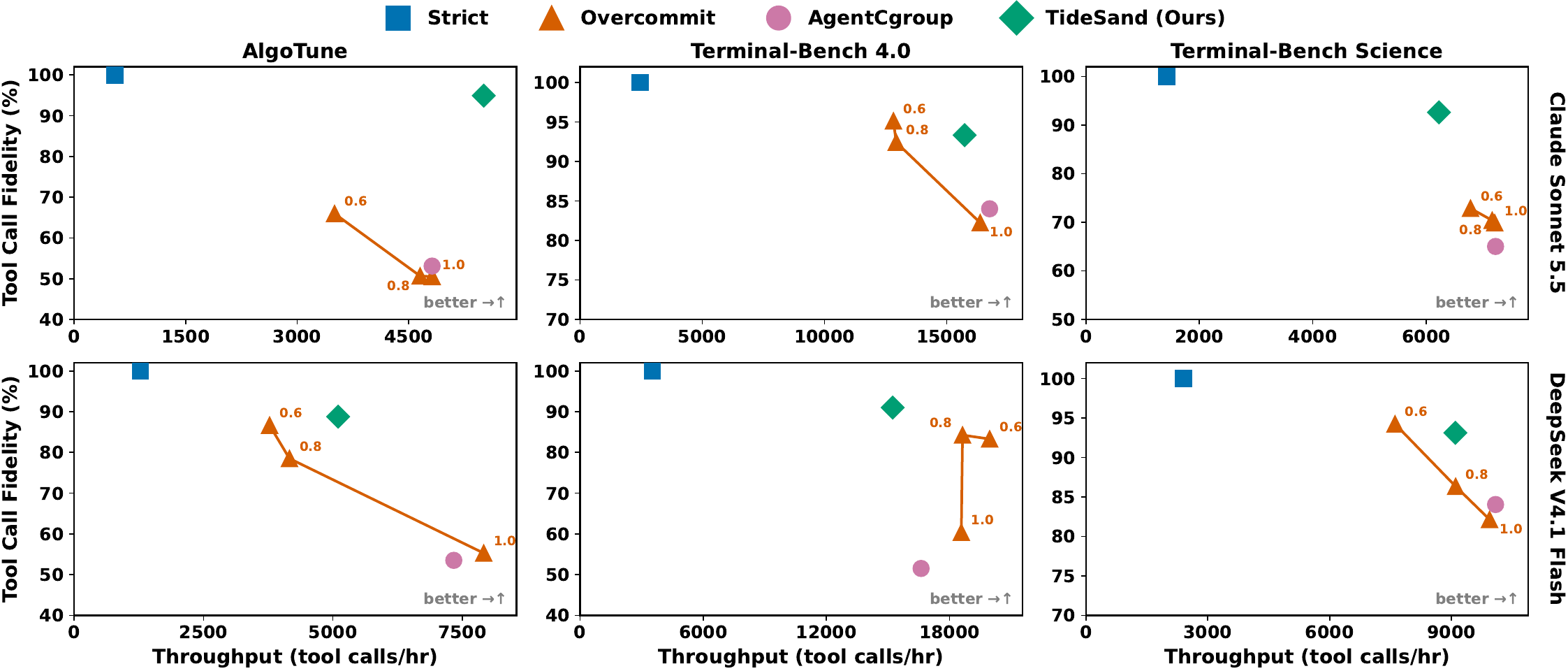}
    \vspace{-15pt}
    \tightcaption{\name achieves better tradeoff on throughput and fidelity across six model-workload pairs. Higher throughput and fidelity is better.}
    \vspace{-5pt}
    \label{fig:eval-e2e}
\end{figure*}

\section{Evaluations}
\label{sec:eval}
\subsection{Experiment Setup}
\mypara{Models and agent}
We use two models: Claude Sonnet 5.5 and DeepSeek V4.1 Flash.
The harness is Claude Code (versions 2.1.300--2.1.309), which runs unmodified inside each sandbox with the model reached through the \name proxy.
Each task was run with the agent's default settings for the benchmark (effort \emph{max} for AlgoTune and Terminal-Bench Science, \emph{medium} for Terminal-Bench 4.0 on Sonnet), and every experiment replays these recordings as described in \S\ref{sec:implementation} to compare on identical agent behavior.

\mypara{Hardware}
Each experiment runs on an AWS \texttt{c7g.metal} bare-metal instance: 64 Arm Neoverse-V1 cores (Graviton3), 128\,GB of DRAM, \verb|gp3| storage, Ubuntu with Linux 6.17.
The sandboxes run in a cgroup of 56 cores and 112\,GiB; the remaining 8 cores run the replay driver and the daemon.
A 64\,GB swap file backs speculative swapping and the memory overcommitment of the baselines.

\mypara{Workloads}
We use three agent benchmarks with resource-intensive tools.
\begin{packeditemize}
    \item AlgoTune~\cite{algotune}: the agent optimizes the runtime of an algorithm while preserving correctness, profiling candidate implementations in the sandbox; 50 tasks, each with 8 CPUs and 16\,GB.
    \item Terminal-Bench 4.0~\cite{terminalbench,terminalbench4}: terminal tasks verified by task-specific tests; 50 recorded tasks , with 2--16 CPUs and 4--16\,GB each.
    \item Terminal-Bench Science~\cite{terminalbenchscience}: data-analysis and simulation tasks derived from published research; a random 50 of the 62 recorded tasks, with 1--4 CPUs and 2--16\,GB each.
\end{packeditemize}
Every task is capped at two hours of agent time.
To load the host with enough load, the driver offers the recorded tasks at a fixed arrival rate and repeats each recording: AlgoTune at $\times$2 (100 sandboxes, one every 15\,s), Terminal-Bench at $\times$8 (400 sandboxes, one every 3\,s), and Terminal-Bench Science at $\times$2 (100 sandboxes, one every 15\,s).

\mypara{Baselines} We compare \name with three policies implemented in the same daemon:
\begin{packeditemize}
    \item Strict reservation: a sandbox is admitted only when its configured CPU and memory fit in the physical capacity.
    \item Overcommit-$\theta$: a sandbox is admitted whenever the host's measured CPU and memory use are both below a fraction $\theta$ of capacity, with $\theta\in\{0.6, 0.8, 1.0\}$; once admitted, sandboxes run without further control. This is similar to the threshold-based overcommit of DeepSeek's DSec~\cite{DSec}.
    \item AgentCgroup~\cite{AgentCgroup}: the tool-call resource controller of AgentCgroup, ported as its cgroup-v2 control loop (Appendix~\ref{app:agentcgroup}). Each tool call is classed by a hint and given a memory allowance and a CPU weight; under memory pressure, high-class sandboxes are protected and the rest are throttled.
\end{packeditemize}

\mypara{Metrics} We evaluate on the  following metrics:
\begin{packeditemize}
    \item Throughput: tool calls completed per host-hour. We measure until 80\% of the run's tool calls have completed to prevent stragglers dominating delay.
    \item Tool-call fidelity: the fraction of tool calls whose delay inflation is less than 10\% of their reference time, which is the slowest of five strict replays of the same recording on strictly reserved host. We only measure fidelity for calls with a reference time of at least one second.
    \item Replay-reward error: the percentage change in tasks' verifier score running in new sandbox system against the strict replay of the same recording. 
\end{packeditemize}
\begin{figure}[!t]
    \centering
    \includegraphics[width=\columnwidth]{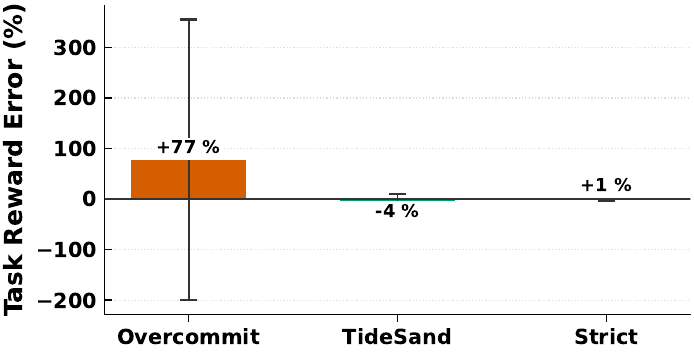}
    \tightcaption{\name reduces the actual reward error on AlgoTune compared to using overcommit policy during actual evaluations.}
    \vspace{-5pt}
    
    \label{fig:eval-task-score}
\end{figure}
\begin{figure}[!t]
    \centering
    \includegraphics[width=\columnwidth]{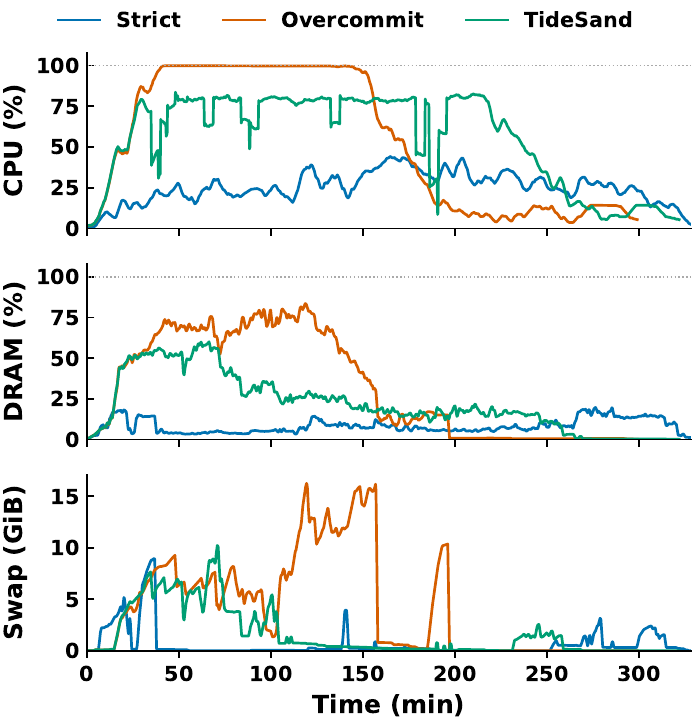}
    \tightcaption{Time series of host CPU, DRAM, and swap use during Terminal-Bench Science runs with Claude Sonnet 5.5 under strict reservation, overcommit-0.8, and \name.}
    \vspace{-5pt}
    \label{fig:eval-utilization}
\end{figure}
\vspace{-5pt}
 \subsection{End-to-End Results}
\label{sec:eval-e2e}

\mypara{Throughput and Fidelity}
Figure~\ref{fig:eval-e2e} shows the throughput and fidelity of the four policies on the six model-workload pairs.
Strict keeps fidelity but runs at a significantly lower throughput.
On the other hand, \name achieves higher throughput while maintaining similar fidelity with strict, achieving better tradeoff compared with baselines.
AgentCgroup's performance is close to overcommit-1.0 on every workload: its throughput stays within 11\% of overcommit-1.0's, and its fidelity ranges from 52\% to 84\%, 9--42\% below \name's.
Notice that sometimes Overcommit achieves better throughput than \name despite having a worse tradeoff curve.
We did not extend \name to that tradeoff point because in practice low fidelity (e.g., under 80\%) could lead to a failed agent.

\begin{figure}[!t]
    \centering
    \includegraphics[width=\columnwidth]{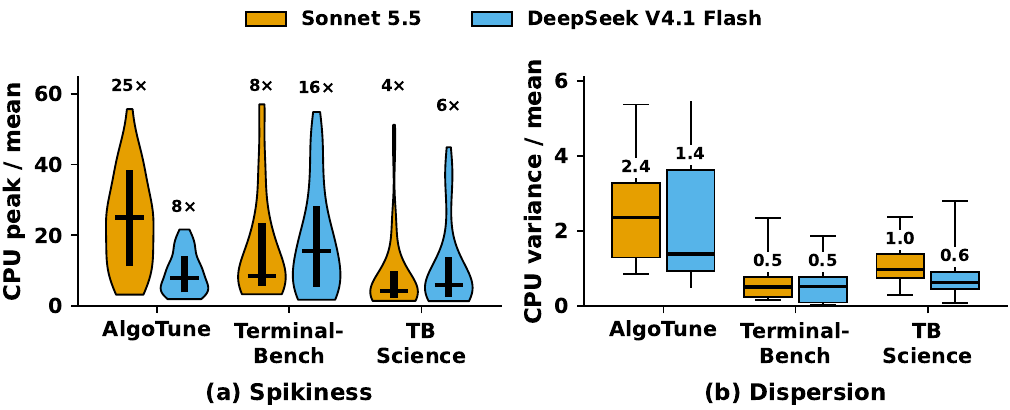}
    \tightcaption{CPU burstiness of each workload under strict reservation (a) each sandbox's peak-to-mean CPU; (b) the variance-to-mean ratio of each sandbox's CPU, in cores. AlgoTune Sonnet has the highest spikiness and dispersion.}
    \vspace{-5pt}
    \label{fig:eval-burstiness}
\end{figure}

\mypara{Workload Burstiness}
\name's advantage is largest on AlgoTune with Sonnet, and Figure~\ref{fig:eval-burstiness} shows why: its sandboxes are the spikiest (peak 25$\times$ the mean CPU, against 4--16$\times$ in other workloads) and the most dispersed.
Dispersion, the variance of a sandbox's CPU use over its mean, measures how much the CPU demand fluctuates for each sandbox.
The baselines do not predict bursts when admitting sandboxes, so when several burst at once they saturate the host.
\name admits a tool call only when it fits the measured headroom, so the more dispersed the workload, the more it gains over baselines.

\mypara{Task-Score Error}
We measure the significance of \name's fidelity increase in Figure~\ref{fig:eval-task-score}.
We replay the same recorded trace of AlgoTune under Sonnet of each task and re-score the identical final solution with the benchmark's verifier under different sandbox policies.
Against the strict replay, overcommit changes the reward by $+77\%$ on average with a standard deviation of 278\% over the 32 scored tasks, 
while \name changes it by $-4\% \pm 14\%$, significantly reducing the error.
 Notice that a second strict run differs from the first by $+1\% \pm 5\%$, which is almost indistinguishable from \name's error.
\mypara{Resource Utilization}
Figure~\ref{fig:eval-utilization} shows how the host's CPU, DRAM, and swap use changes over time during the Terminal-Bench Science's Sonnet runs.
Strict reservation keeps the host at 23\% CPU on average for six hours and never exceeds a fifth of its DRAM.
Overcommit-0.8 admits the workload within its first hour and holds the host at 100\% CPU for the next hour while resident memory and swap climbs.
This damages the fidelity of the task running during this period.
\name runs the same workload at below 80\% CPU and 60\% DRAM utilization, resulting in tool call fidelity preservation.

\begin{figure}[!t]
    \centering
    \includegraphics[width=0.9\columnwidth]{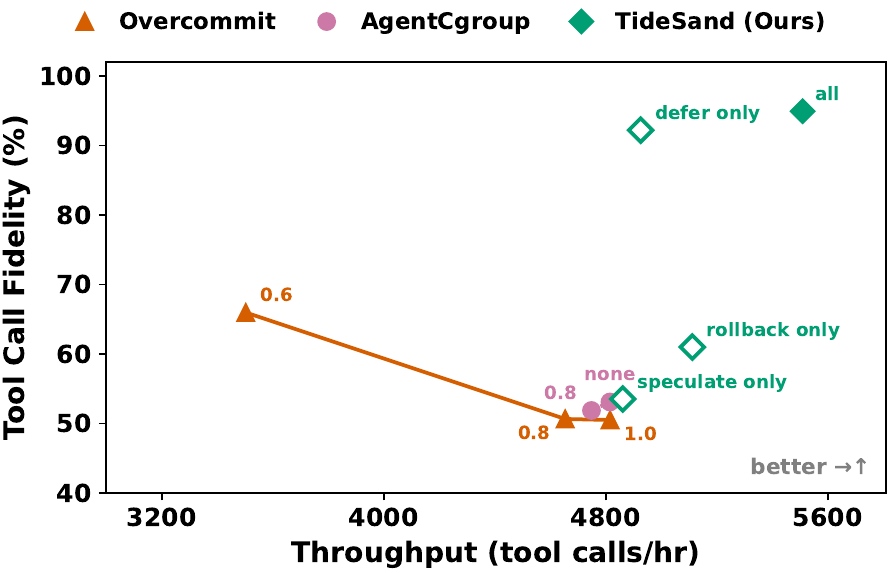}
    \tightcaption{Ablation of \name on AlgoTune ($\times$2). We compare individual mechanisms of \name against the full policy.}
    \label{fig:eval-mechanisms}
\end{figure}
\vspace{-5pt}
\begin{figure}[!t]
    \centering
    \includegraphics[width=\columnwidth]{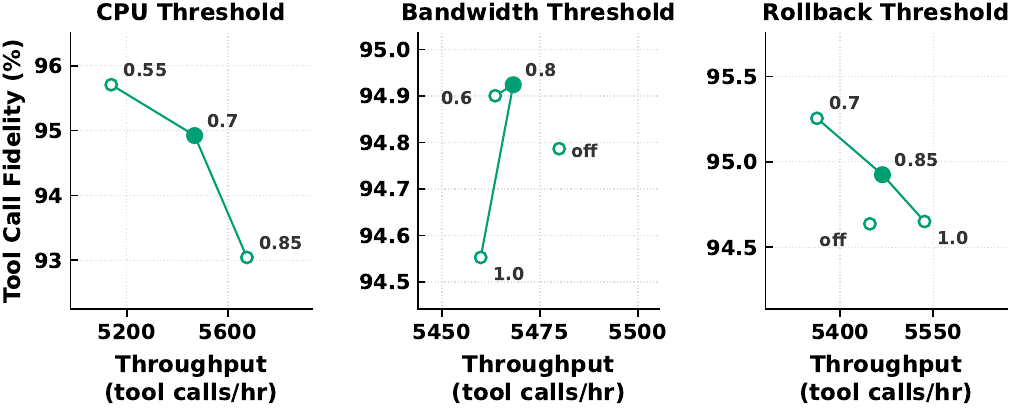}
    \vspace{-15pt}
    \tightcaption{\name remains robust across different hyperparameter setups on AlgoTune dataset running under Sonnet 5.5.}
    \label{fig:eval-ablation}
\end{figure}
\subsection{Extended Evaluation}
\label{sec:eval-extended}
\vspace{5pt}
\begin{figure}[!t]
    \centering
    \includegraphics[width=\columnwidth]{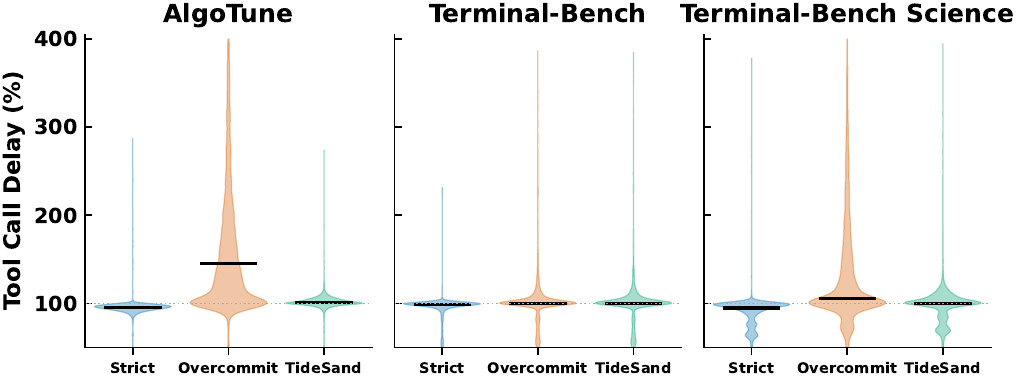}
    \vspace{-15pt}
    \tightcaption{\name keeps tool-call distributions similar to Strict across workloads}
    \label{fig:eval-tool-dist}
\end{figure}

\begin{figure}[!t]
    \centering
    \includegraphics[width=\columnwidth]{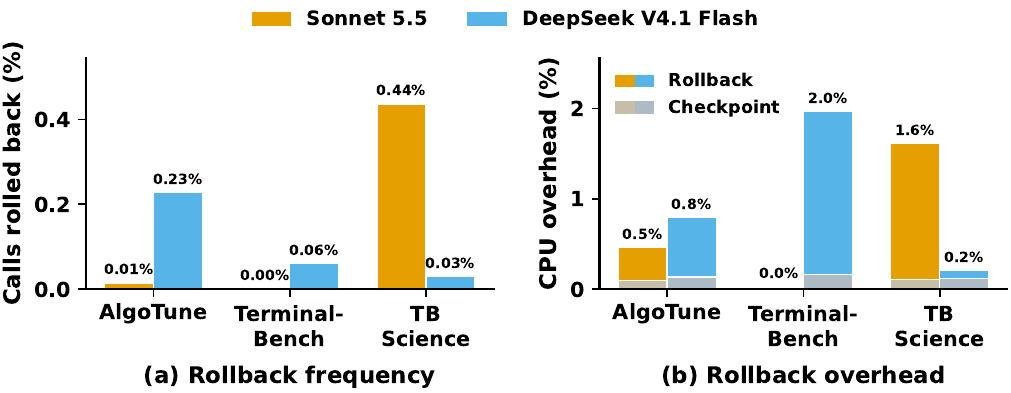}
    \tightcaption{\name rolls back few tool calls, with CPU overhead under 2\% across workloads.}
    \label{fig:eval-micro}
    \vspace{-5pt}
\end{figure}
\vspace{-5pt}

\mypara{Tool-Call Distributions}
Figure~\ref{fig:eval-tool-dist} shows each tool call's execution time relative to its reference for the three Sonnet workloads.
Strict concentrates tightly around the reference, while overcommit develops a long tail of heavily slowed calls.
\name stays close to strict and removes most of this tail.
Terminal-Bench shows the smallest gap because its calls are short and rarely saturate the host.

\mypara{Mechanism Ablation}
Figure~\ref{fig:eval-mechanisms} runs \name with one mechanism at a time on AlgoTune with Sonnet.
Each mechanism alone moves above the overcommit curve, with defer only lands close to full policy 
but sacrifices throughput.  Speculative swapping and rollback are slightly above overcommit for their ability to clear more spaces to reduce contention.
However, no single mechanism is able to achieve the best tradeoff as \name does.

\mypara{Parameter Sensitivity}
Figure~\ref{fig:eval-ablation} varies one parameter at a time from the default configuration on AlgoTune with Sonnet.
Across all these settings, fidelity stays between 93\% and 96\%, and no setting causes a sharp drop in either metric.
Changing these parameters only slightly shifts \name along its throughput-fidelity tradeoff curve, 
but does not dampen the design's superiority over baselines.

\mypara{Microbenchmarking}
Figure~\ref{fig:eval-micro} measures how often selective rollback acts and what it costs across the six end-to-end runs.
Only 0.6--2.9\% of calls are checkpointed, and rollbacks are rare: at most 0.44\% of calls, on Terminal-Bench Science with Sonnet.
Counting both checkpoint freezes and discarded execution, the CPU overhead stays below 2\% on every workload.
This shows that \name rarely falls back to rollback and the cost is minimal. 
\vspace{-5pt}
\section{Related Work}
\label{sec:related-work}
\mypara{LLM Inference}
LLM serving systems improve inference through memory management and attention kernels~\cite{vllm,sglang,flashinfer},
prefill-decode disaggregation~\cite{distserve,mooncake}, and KV-cache compression, reuse, and offloading~\cite{cachegen,cacheblend,cachedattention,lmcache}.
For agent workloads, Agentix~\cite{agentix} schedules calls using program-level history,
Continuum~\cite{continuum} retains KV caches during tool calls with bounded time-to-live,
and MORI~\cite{mori} uses relative idleness to place caches across GPU and CPU memory.
ThunderAgent~\cite{thunderagent} coordinates program-aware inference scheduling with tool-resource management.
\name addresses the complementary challenge of sharing CPU and memory for sandboxes.

\mypara{Serverless Container and Sandbox}
Serverless systems reduce environment startup and memory costs through caching, prefetching, and reclamation.
Catalyzer~\cite{catalyzer} reuses initialized state for fast sandbox startup,
and MITOSIS~\cite{mitosis} enables RDMA-based remote fork without provisioning cached instances on each host.
RainbowCake~\cite{rainbowcake} shares container layers, REAP~\cite{reap} prefetches snapshot working sets,
and FaaSMem~\cite{faasmem} and Squeezy~\cite{squeezy} improve memory efficiency through offloading and reclamation.
Jiagu~\cite{jiagu} combines performance-aware scheduling and staged scaling to improve utilization while maintaining QoS.
\name complements these mechanisms by exploiting agent specific improvements by efficient sharing while controlling contention for fidelity to preserve autoresearch agent measurements.

\mypara{Commercial Sandbox}
We describe the oversubscription policies of two leading commercial sandbox providers:
E2B supports memory oversubscription through userfaultfd-based lazy memory loading, 
allocating physical memory only as sandbox pages are accessed~\cite{e2bcompose,e2barchitecture}.
Modal supports CPU oversubscription by separating guaranteed CPU reservations from maximum usage limits,
 allowing sandboxes to burst beyond their reserved capacity when spare resources are available~\cite{modalsandbox,modalresources}.
However, 
both policies are designed to improve execution time rather than fidelity.
Our focus is preserving that criterion for autoresearch measurements while sharing idle  resources.

\vspace{-5pt}
\section{Future Work}
\label{sec:discussion}

\mypara{Extension to More Sandbox Types}
Extending \name to GPU or TPU sandboxes requires tracking and reclaiming accelerator state alongside CPU memory. Tools may load model weights, allocate device buffers, or launch kernels, so restoring a sandbox must preserve consistency between host and device state while weighing transfer costs against expected idle intervals. While utilities like NVIDIA's CUDA checkpoint~\cite{cudacheckpoint} support saving and restoring GPU state alongside CPU checkpoints, arbitrary accelerator workloads and other backends (such as TPUs) require assessing backend-specific runtime and driver capabilities.

\mypara{Co-design with Inference System}
Sandbox and inference systems can coordinate in both directions: 
inference progress can guide sandbox restoration, while tool execution signals can inform KV-cache retention. 
For example, Continuum uses tool-aware time-to-live policies to keep an agent's KV cache in GPU memory across short tool calls~\cite{continuum}. 
A joint scheduler could refine these decisions using live signals, such as evicting KV cache when sandbox server is already under contention, or reclaiming sandbox memory during extended inference waits.

\mypara{Better Mechanism Support}
Our main contribution in \name is outlining three key ideas for obtaining performance accurate sandboxes. However, the mechanisms that implement these ideas, can be made more efficient and left as future work. For example, we found no good preexisting read-to-use methods for implementing memory bandwidth isolation between sandboxes, and this made perfect tool-call fidelity hard to obtain. Checkpoint and restore overhead can also be further reduced with more efficient mechanisms such as \cite{DeltaBox}.

\vspace{-5pt}
\section{Conclusion}
\label{sec:conclusion}

Agent sandboxes introduce new opportunities for sandbox management: bursty tool execution allows resource multiplexing for more efficient resource utilization.
However, resource contention can distort the measurements that guide subsequent agent decisions, compromising tool-call fidelity and task quality.
We present \name, which combines agent-aware scheduling, speculative swapping, and selective rollback to obtain performance accurate sandbox multiplexing.
Across six model-workload pairs, our evaluation shows that \name achieves 3.8--10$\times$ the throughput of strict reservation while preserving tool-call fidelity, 7--44\% higher fidelity than overcommitting at similar throughput, and reduces the error in the agent's task reward from 77\% to 4\%.
We hope this work motivates the development of more performance-accurate sandbox systems for LLM agents that operate with high resource utilization.

\bibliography{example_paper}
\bibliographystyle{mlsys2025}

\section{AI Usage Disclosure}
We mainly used coding agents to help generate code. We also used AI tools to proofread, correct grammar, and refine sentences for writing.
\onecolumn
\appendix

\section{LLM Price Data}
\begin{table}[!ht]
\centering
\scriptsize
\caption{Models scoring at least 50\% on Terminal-Bench 2.1, as plotted in Figure~\ref{fig:llm_vs_sandbox_prices}(a).
Scores are the best independent result recorded in our data: T is the official tbench.ai leaderboard~\cite{tbleaderboard}, V is Vals AI (Terminus 2 harness)~\cite{valstb21}, and A is Artificial Analysis~\cite{aatb21}.
Prices are listed input/output rates in USD per 1M tokens compiled in October 2026; blended weights input and output 3:1.
Bold rows set a new lowest blended price.}
\label{tab:llm_prices}
\begin{tabular}{@{}lllccc rrr@{}}
\toprule
Model & Developer & Release & Open & TB2.1 (\%) & Src & Input & Output & Blended \\
\midrule
\textbf{Gemini 3 Pro} & Google & 2025-11-18 &  & 73.9 & T & 2.00 & 12.00 & 4.500 \\
\textbf{Gemini 3 Flash} & Google & 2025-12-17 &  & 53.9 & V & 0.50 & 3.00 & 1.125 \\
Claude Sonnet 4.6 & Anthropic & 2026-02-17 &  & 57.3 & V & 3.00 & 15.00 & 6.000 \\
Gemini 3.1 Pro & Google & 2026-02-19 &  & 70.8 & V & 2.00 & 12.00 & 4.500 \\
GPT-5.4 mini & OpenAI & 2026-03-17 &  & 54.7 & V & 0.75 & 4.50 & 1.688 \\
GLM-5.1 & Zhipu & 2026-03-27 & \checkmark & 58.6 & T & 1.00 & 3.20 & 1.550 \\
Qwen3.6 Plus & Alibaba & 2026-03-31 &  & 53.2 & V & 0.50 & 3.00 & 1.125 \\
Claude Opus 4.7 & Anthropic & 2026-04-16 &  & 68.9 & T & 5.00 & 25.00 & 10.000 \\
Kimi K2.6 & Moonshot & 2026-04-20 & \checkmark & 53.6 & V & 0.95 & 4.00 & 1.712 \\
\textbf{MiMo V2.5} & Xiaomi & 2026-04-22 & \checkmark & 60.7 & V & 0.14 & 0.28 & 0.175 \\
MiMo V2.5 Pro & Xiaomi & 2026-04-22 & \checkmark & 57.3 & V & 0.435 & 0.87 & 0.544 \\
GPT-5.5 & OpenAI & 2026-04-23 &  & 83.2 & T & 5.00 & 30.00 & 11.250 \\
DeepSeek V4 & DeepSeek & 2026-04-24 & \checkmark & 50.2 & V & 1.32 & 3.96 & 1.980 \\
Composer 2.5 & Cursor & 2026-05-18 &  & 58.4 & V & 0.50 & 2.50 & 1.000 \\
Gemini 3.5 Flash & Google & 2026-05-19 &  & 74.2 & V & 1.50 & 9.00 & 3.375 \\
Qwen3.7 Max & Alibaba & 2026-05-19 &  & 61.0 & V & 2.50 & 7.50 & 3.750 \\
Claude Opus 4.8 & Anthropic & 2026-05-28 &  & 78.9 & T & 5.00 & 25.00 & 10.000 \\
MiniMax-M3 & MiniMax & 2026-06-01 & \checkmark & 65.2 & A & 0.60 & 2.40 & 1.050 \\
Qwen3.7 Plus & Alibaba & 2026-06-02 &  & 52.8 & V & 0.40 & 1.60 & 0.700 \\
Nemotron 3 Ultra & NVIDIA & 2026-06-03 & \checkmark & 50.9 & V & 0.60 & 2.40 & 1.050 \\
Claude Fable 5 & Anthropic & 2026-06-09 &  & 83.8 & T & 10.00 & 50.00 & 20.000 \\
Kimi K2.7 Code & Moonshot & 2026-06-12 & \checkmark & 67.0 & V & 0.95 & 4.00 & 1.712 \\
GLM-5.2 & Zhipu & 2026-06-16 & \checkmark & 67.8 & V & 1.40 & 4.40 & 2.150 \\
Claude Sonnet 5 & Anthropic & 2026-06-30 &  & 74.6 & T & 2.00 & 10.00 & 4.000 \\
Grok 4.5 & xAI & 2026-07-08 &  & 79.3 & T & 2.00 & 6.00 & 3.000 \\
GPT-5.6 Luna & OpenAI & 2026-07-09 &  & 79.0 & V & 0.20 & 1.20 & 0.450 \\
GPT-5.6 Sol & OpenAI & 2026-07-09 &  & 89.5 & A & 4.00 & 20.00 & 8.000 \\
GPT-5.6 Terra & OpenAI & 2026-07-09 &  & 78.4 & T & 2.00 & 12.00 & 4.500 \\
Muse Spark 1.1 & Meta & 2026-07-09 &  & 76.2 & T & 1.25 & 4.25 & 2.000 \\
Inkling & Thinking Machines & 2026-07-15 & \checkmark & 55.1 & A & 1.00 & 4.05 & 1.762 \\
Kimi K3 & Moonshot & 2026-07-16 & \checkmark & 85.0 & A & 3.00 & 15.00 & 6.000 \\
Gemini 3.5 Flash-Lite & Google & 2026-07-21 &  & 50.2 & V & 0.30 & 2.50 & 0.850 \\
Gemini 3.6 Flash & Google & 2026-07-21 &  & 73.8 & V & 1.50 & 7.50 & 3.000 \\
\textbf{Ling 3.0 Flash} & InclusionAI & 2026-07-23 & \checkmark & 50.2 & V & 0.075 & 0.22 & 0.111 \\
Claude Opus 5 & Anthropic & 2026-07-24 &  & 84.6 & V & 5.00 & 25.00 & 10.000 \\
Inkling Small & Thinking Machines & 2026-07-30 & \checkmark & 55.1 & V & 0.30 & 1.20 & 0.525 \\
DeepSeek V4 Flash 0731 & DeepSeek & 2026-07-31 & \checkmark & 67.0 & V & 0.44 & 1.32 & 0.660 \\
Qwen3.8 Max & Alibaba & 2026-08-02 &  & 67.4 & V & 2.00 & 6.00 & 3.000 \\
Muse Spark 1.2 & Meta & 2026-08-05 &  & 69.7 & V & 1.25 & 4.25 & 2.000 \\
Grok 4.6 & xAI & 2026-08-12 &  & 78.3 & V & 2.00 & 6.00 & 3.000 \\
DeepSeek V4 Pro 0813 & DeepSeek & 2026-08-13 & \checkmark & 54.7 & V & 1.32 & 3.96 & 1.980 \\
Gemini 3.7 Flash & Google & 2026-08-13 &  & 77.5 & V & 1.50 & 7.50 & 3.000 \\
Qwen3.8 27B & Alibaba & 2026-08-14 & \checkmark & 79.8 & A & 0.50 & 3.00 & 1.125 \\
GLM-5.3 & Zhipu & 2026-08-18 & \checkmark & 83.9 & A & 1.40 & 4.40 & 2.150 \\
GLM-5.3-Flash & Zhipu & 2026-08-26 & \checkmark & 84.3 & A & 0.075 & 0.25 & 0.119 \\
Hy4 Preview & Tencent & 2026-08-27 & \checkmark & 55.1 & V & 0.834 & 2.501 & 1.251 \\
Claude Fable 5.1 & Anthropic & 2026-09-01 &  & 91.4 & A & 10.00 & 50.00 & 20.000 \\
Gemini 3.8 Flash & Google & 2026-09-02 &  & 87.6 & A & 1.50 & 7.50 & 3.000 \\
Muse Spark 1.3 & Meta & 2026-09-02 &  & 79.0 & V & 1.25 & 4.25 & 2.000 \\
Qwen3.8 Max 0902 & Alibaba & 2026-09-02 &  & 88.8 & A & 2.00 & 6.00 & 3.000 \\
GPT-6 Astra & OpenAI & 2026-09-03 &  & 89.9 & A & 10.00 & 50.00 & 20.000 \\
DeepSeek V4.1 Flash & DeepSeek & 2026-09-09 & \checkmark & 74.5 & V & 0.30 & 1.20 & 0.525 \\
Grok 4.7 & xAI & 2026-09-21 &  & 73.4 & V & 2.00 & 6.00 & 3.000 \\
MiMo V2.6 Flash & Xiaomi & 2026-09-21 & \checkmark & 76.4 & V & 0.14 & 0.28 & 0.175 \\
MiMo V2.6 Pro & Xiaomi & 2026-09-21 & \checkmark & 67.8 & V & 0.435 & 0.87 & 0.544 \\
Claude Opus 5.5 & Anthropic & 2026-09-22 &  & 87.6 & V & 4.00 & 20.00 & 8.000 \\
GPT-6 Luna & OpenAI & 2026-09-22 &  & 73.0 & V & 0.10 & 0.50 & 0.200 \\
GPT-6 Sol & OpenAI & 2026-09-22 &  & 83.2 & V & 2.00 & 10.00 & 4.000 \\
Claude Sonnet 5.5 & Anthropic & 2026-09-28 &  & 83.2 & V & 2.00 & 10.00 & 4.000 \\
Gemini 4 Argon & Google & 2026-09-30 &  & 57.7 & V & 4.00 & 20.00 & 8.000 \\
Muse Spark Open & Meta & 2026-10-15 & \checkmark & 76.4 & A & 1.25 & 4.25 & 2.000 \\
\bottomrule
\end{tabular}
\end{table}

\section{AgentCgroup Baseline}
\label{app:agentcgroup}

The AgentCgroup baseline in \S\ref{sec:eval} is a port of the tool-call resource controller of AgentCgroup~\cite{AgentCgroup}, run as its cgroup-v2 control loop inside the same daemon as the other policies.

Each tool call, delimited by the sandbox's model requests, opens a resource domain.
A hint derived from the call (file tools and light shell commands low, package fetches medium, everything else high) sets the sandbox's \texttt{memory.high} to the call's allowance over the sandbox's memory at the call's start (256\,MiB, 1\,GiB, or the limit) and its CPU weight to its class's (150 high, 50 low).
Memory pressure, from host memory use above 85\%, memory PSI above 10\%, or a high-class call stalling on memory, opens one-second protection windows in which high-class sandboxes are protected with \texttt{memory.low} and low-class sandboxes are squeezed to their current use plus 256\,MiB and frozen after three seconds over their cap.
The controller's per-call child cgroups cannot be created on the container's exec path, so the domain is the sandbox's cgroup, and the agent's response to the controller's throttling feedback, re-issuing the call with a higher hint, is replayed as a promotion of the hint after 30 seconds of throttling.


\end{document}